\documentclass{article}
\usepackage{cite,url,amsfonts,amsmath}
\usepackage{algorithm}
\usepackage{algpseudocode}
\usepackage{mathtools}
\usepackage{graphicx}
\usepackage{color}
\usepackage{hyperref}
\DeclareMathOperator*{\argmin}{arg\,min}

\begin{document}
\title{Artistic Curve Steganography Carried by Musical Audio}
%
%
\author{Christopher J Tralie, Ursinus College, ctralie@alumni.princeton.edu}
%
%
%
\maketitle              
\begin{abstract}

  In this work, we create artistic closed loop curves that trace out images and 3D shapes, which we then hide in musical audio as a form of steganography.  We use traveling salesperson art to create artistic plane loops to trace out image contours, and we use Hamiltonian cycles on triangle meshes to create artistic space loops that fill out 3D surfaces. Our embedding scheme is designed to faithfully preserve the geometry of these loops after lossy compression, while keeping their presence undetectable to the audio listener. To accomplish this, we hide each dimension of the curve in a different frequency, and we perturb a sliding window sum of the magnitude of that frequency to best match the target curve at that dimension, while hiding scale information in that frequency's phase.  In the process, we exploit geometric properties of the curves to help to more effectively hide and recover them.  Our scheme is simple and encoding happens efficiently with a nonnegative least squares framework, while decoding is trivial.  We validate our technique quantitatively on large datasets of images and audio, and we show results of a crowd sourced listening test that validate that the hidden information is indeed unobtrusive.

\end{abstract}
\section{Introduction / Background}

Steganography is the process of hiding one data stream ``in plain sight'' within another data stream known as a ``carrier.''  In this work, we are interested in using audio as a carrier to store images, which has been relatively unexplored in the academic literature (Section~\ref{sec:priorwork}).  As with any stegonographic technique, one can use this as a general means of covert communication, artists can also use it to hide images and to watermark their tunes.  One such example occurred when Aphex Twin embedded a face image inside the spectrogram of their ``Windowlicker'' song \cite{mathews2004music}, though the technique yields very loud, inharmonic, scratchy sounds.  By contrast, we seek an embedding of the image that is undetectable.  Like Aphex Twin, however, {\em we would like the image to survive lossy audio compression} so that people can communicate their images via social media, which makes the problem significantly more challenging.

To make our audio steganography system as robust and as usable as possible, we have the following design goals:
\begin{enumerate}
    \item \label{goal:imperceptible} The hidden data should be audibly imperceptible
    \item \label{goal:geomquality} The hidden data should be faithfully preserved {\em under lossy compression}
    \item \label{goal:misalignment} The hidden data should be robust to frame misalignment, or it should be possible to recover a frame alignment without any prior information
    \item \label{goal:metadata} No metadata should be required to retrieve the hidden data; (lossily compressed) audio alone should suffice
    \item \label{goal:partial} It should be possible to recover the data partially from partial audio chunks; that is, we don't need to wait for the entire data stream to recover the signal
\end{enumerate}

Goals ~\ref{goal:imperceptible} and ~\ref{goal:geomquality} are at odds with each other, and satisfying them simultaneously is the biggest challenge of this work.  We formulate an objective function in Section~\ref{sec:formulation} to trade off both of these goals.

Rather than storing images directly in audio, we constrain our problem to hiding artistic curves that trace out images (Section~\ref{sec:tspart}) or 3D surface shapes (Section~\ref{sec:hamiltonian}).  We hide each dimension of our curve in a different frequency, and we use sliding window sums of the frequency magnitudes to smooth them in time, which makes them more robust to compression, as we show in Section~\ref{sec:experiments}.  Furthermore, since we know that the curves only move by a small amount between adjacent samples in time, we can exploit this fact to recover from frame misalignments to satisfy Goal~\ref{goal:misalignment}, as explained in Section~\ref{sec:framealignments}.  In the end, all of the hidden information needed to reconstruct the curves is stored in the magnitudes and phases of frequencies in the audio, so no metadata is needed (Goal~\ref{goal:metadata}).  Finally, the information needed to recover individual curves samples is localized in time, so the curves can be partially decoded from partial audio (Goal~\ref{goal:partial}).

\subsection{Prior Work in Audio Steganography}
\label{sec:priorwork}

Before we proceed to describing the techniques for constructing artistic curves (Section~\ref{sec:tspart}, Section~\ref{sec:hamiltonian}), and to ultimately hiding them in audio (Section~\ref{sec:methods}), we briefly review adjacent work in the area of audio steganography to put our work in context (for more information, review to these survey articles \cite{djebbar_comparative_2012, dutta_overview_2020}).

As is the case for most other steganographic carriers, the simplest techniques for audio steganography rely on changing the least significant bit of uncompressed encodings (e.g. \cite{cvejic_wavelet_2002}).  Other early works rely on the ``frequency masking'' property of the human auditory system.  Binary data is stored by controlling the relative amplitude of two frequencies that are masked in this way \cite{gopalan_unified_2009, gopalan2004audio}.

Beyond amplitude perturbations, it is also possible to keep the amplitudes fixed and to change the phases \cite{xiaoxiao_dong_data_2004, yun2009acoustic}; the technique of \cite{malik_robust_2007} was particularly successful at this by manipulating the poles of allpass filters to encode binary data.  One can also hide binary data by adding and manipulating echoes of a signal \cite{gruhl1996echo}, which is more robust than other techniques to lossy compression since it is akin to impulse responses of natural environments; however, the bit rate is quite low (on the order of dozens of bits per second).  For another approach to compressed audio, some techniques are able to use the properties of mp3 files directly to hide information \cite{qiao_steganalysis_2009,atoum2013exploring}.  Other techniques adapt wireless transmissions schemes, such as and on-off keying (OOK) \cite{madhavapeddy_audio_2005} (at a low bit rate of 8bps) and orthogonal frequency-division multiplexing (OFDM) \cite{eichelberger_receiving_2019} (at a higher bit rate of closer to 1kbps), to the audio domain.

Most of the hidden messages in the audio steganography literature are in binary format, but a few recent works have focused on hiding images pixels using deep learning to train a ``hiding network'' and a ``reveal network'' in tandem to encode an image in an audio stream and then to decode its addition into the carrier, respectively, while maximizing quality of the decoded image and the encoded audio \cite{cui_multi-stage_2021, geleta_pixinwav_2021, takahashi_source_2022, domenech2022hiding}.  Not only do we also attempt to hide images, but we have a similar perspective that rather than attempting to extract binary sequences exactly, we can tolerate progressive noise in the reconstructions.  Challenges of deep-learning based approaches include the need to extensively train on examples, the difficulty of including a loss term that can model the effect of lossy audio compression, and the difficulty of training to be robust to frame misalignment (Goal~\ref{goal:misalignment}).  We sidestep these challenges in our work by using a simple model based on hiding in frequencies.

\subsection{Traveling Salesperson Art}
\label{sec:tspart}

\begin{figure}
  \centering
  \includegraphics[width=\columnwidth]{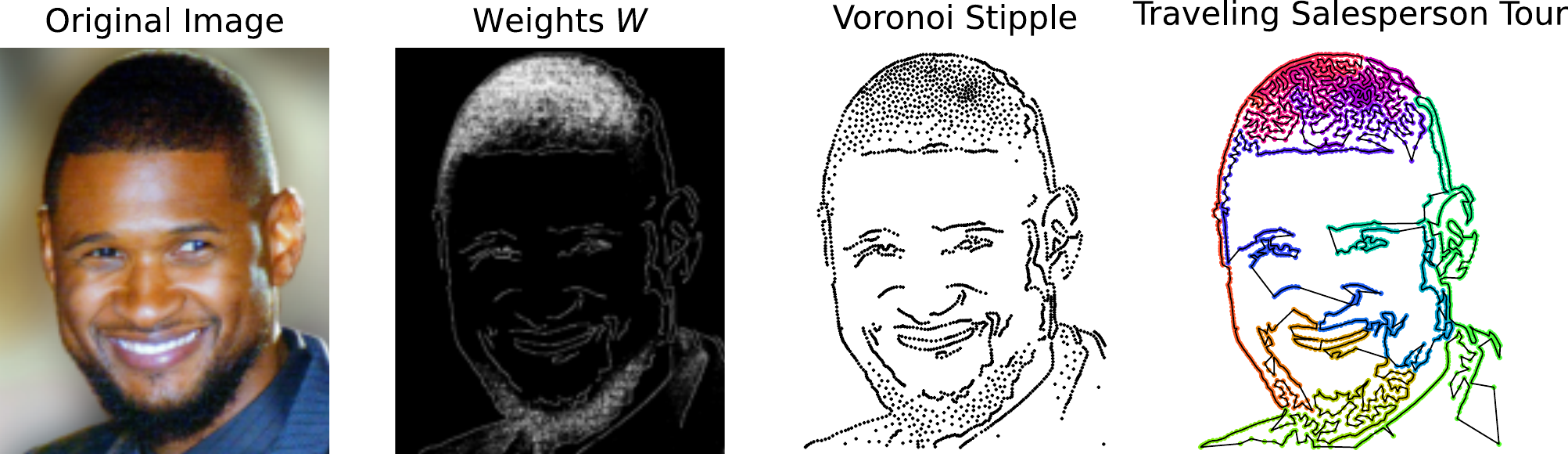}
  \caption{Our modified pipeline for creating TSP art.  Color on the tour indicates phase along the loop.}
  \label{fig:TSPTour}
\end{figure}

To devise 2D curves that stand in for images, we use Traveling Salesperson (TSP) art \cite{bosch2004continuous, kaplan2005tsp,bosch2008connecting}, which is an automated artistic line drawing pipeline which computes a {\em simple closed loop} to approximate a source image.  ``Simple'' in this context does not imply a lack of complexity; rather, it means that a curve does not intersect itself (see, for example, the Jordan Curve Theorem \cite{bosch2009jordan}).  To construct such a curve, we first place a collection of dots in the plane in a ``stipple pattern'' to approximate brightness in the image (e.g. more dots concentrate in darker regions), and then connect them in a loop via an approximate traveling salesperson (TSP) tour.  Figure~\ref{fig:TSPTour} shows an example.  We follow the TSP Art technique of \cite{kaplan2005tsp}, with a few modifications, as described below.

\begin{figure}
  \begin{minipage}[c]{0.4\textwidth}
    \caption{
      Since mp3 compression introduces noise to our embedded curves, we pre-smooth them using one iteration of curvature-shortening flow at $\sigma=1$, which smooths the curves without introducing any crossings.
    } \label{fig:CurvatureShortening}
  \end{minipage}
  \begin{minipage}[c]{0.6\textwidth}
    \includegraphics[width=\columnwidth]{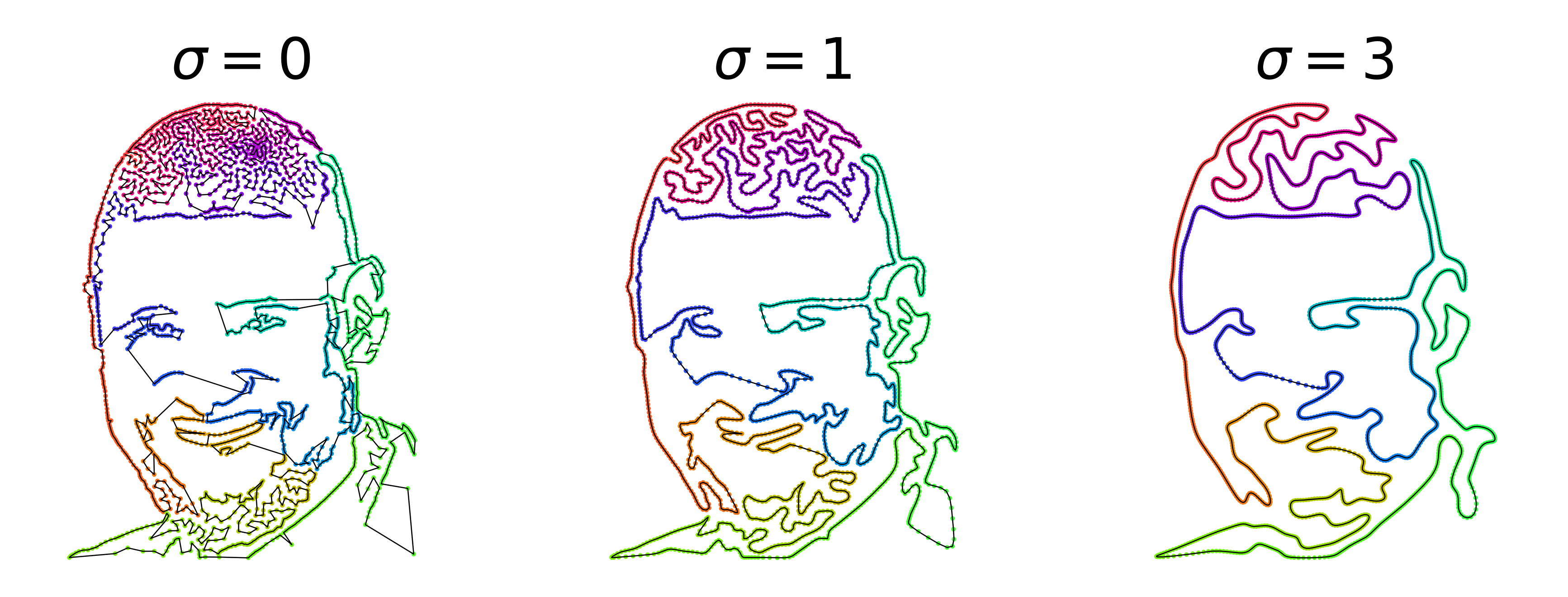}
  \end{minipage}\hfill
\end{figure}

The first step in TSP art is to generate a stipple pattern $X$ to best approximate a grayscale image $G$.  Following the authors of \cite{kaplan2005tsp}, we use Secord's technique for Voronoi stippling \cite{secord2002weighted}, which takes an initial sample of points according to some density weights $W$ and then repeatedly moves them closer the weighted centroid of their Voronoi regions (an instance of Lloyd's algorithm) so that they are spread more evenly.  Secord \cite{secord2002weighted} takes the weight $W_{ij}$ at a pixel to be inversely proportional to its brightness, using a weight of $0$ above some brightness threshold $b$.  This leads more dots to concentrate in darker regions; however, the algorithm may fail to sample any dots along important edges between brighter regions (the authors of \cite{li2011structure} also observed this).  To mitigate this, we run a Canny edge detector \cite{canny1986computational} on the original image and set the weight of any pixels along an edge to be 1, so that the final weights promote samples both in darker regions and along edges of any kind.  This addition is particularly helpful for line drawings.  

Once a stipple has been established, the next step in TSP art is to ``connect the dots'' with a closed loop that visits each stipple point exactly once, referred to as a ``tour''.  A well known objective function for a tour that doesn't ``jump too much'' is the total distance traveled, or the sum of all edge lengths, and a tour that achieves the optimum is known as a {\em traveling salesperson (TSP) tour}.  Since the TSP problem is NP-hard, the authors of  \cite{kaplan2005tsp} use the Concorde TSP solver \cite{applegate2001concorde} for an approximate solution. We opt for a simpler technique that first creates a 2-approximation of a TSP from a depth-first traversal through the minimum spanning tree of the stipple dots, which is a already a 2-approximation of the optimal tour.  We then iteratively improve on this tour via a sequence of 2-opt relaxations \cite{johnson1997traveling}; that is, if for some $i > 1, i < j < N$ the distances between the 4 points $X_i, X_{i+1}, X_j$, and $X_{j+1}$ satisfy

  

\begin{equation}
  d(X_i, X_j) + d(X_{i+1}, X_{j+1}) < d(X_i, X_{i+1}) + d(X_j, X_{j+1})
\end{equation}

then it is possible to perform a swap to yield a new tour with a smaller distance.  This amounts to reversing the indices in the tour between index $i+1$ and $j$, inclusive.  We repeat this step as long as such a swap is still possible.  Though this is not guaranteed to yield an optimal TSP tour, it does produce aesthetically pleasing tours which are simple; that is, every crossing is removed.

\subsubsection{Curvature Shortening Flow}




Since mp3 compression introduces noise into our embedded curves, we smooth them before embedding to improve visual quality.  To this end, we apply a numerical version of curvature shortening flow described by Mokhtarian and Mackworth \cite{mokhtarian1992theory} which applies to piecewise linear curves (like our TSP tours).  The technique works by numerically by convolving coordinates of each curve with smoothed versions of Gaussians and their derivatives.  

To approximately smooth a curve via one step of curvature-shortening flow, Mokhtarian and Mackworth \cite{mokhtarian1992theory} show that it suffices to first re-parameterize by the arc length, and then to smooth the curve by convolving with once with a Gaussian\footnote{The beauty of convolving with Gaussians as such is that $\gamma$ does not even have to be differentiable, so this works on our piecewise linear TSP tours.}.  By the Gage-Hamilton-Grayson theorem, simple curves that undergo curvature shortening flow {\em remain simple} and eventually become convex, shrinking to a point under repeated applications of the flow.  Figure~\ref{fig:CurvatureShortening} shows an example of one application of curvature shortening flow for different $\sigma$ values.

\subsection{Hamiltonian Cycles on Watertight Triangle Meshes}
\label{sec:hamiltonian}

\begin{figure}
  \centering
  \includegraphics[width=0.8\columnwidth]{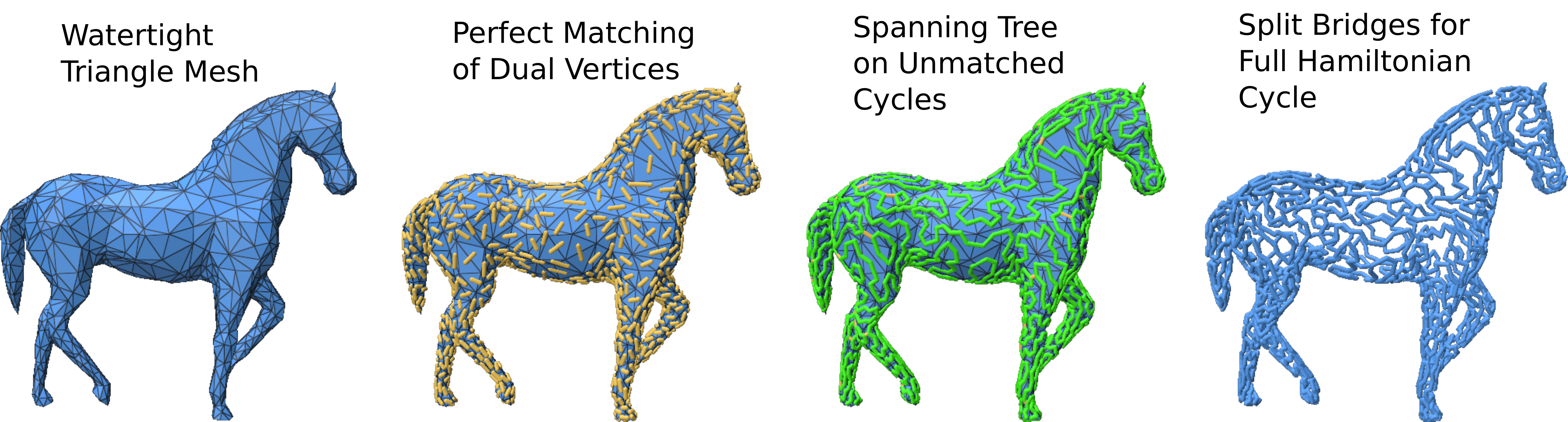}
  \caption{An example of the algorithm of Gopi and Epstein \cite{gopi2004single} on a horse model from the Princeton mesh segmentation benchmark \cite{Chen:2009:ABF}.}
  \label{fig:HamiltonianHorse}
\end{figure}

In addition to 2D TSP tours on stipple patterns on planar images, we also create artistic 3D space loops that fill out the surfaces of 3D shapes following the technique of Gopi and Epstein \cite{gopi2004single}.  They observe that watertight triangle meshes (those with no boundary) have a dual graph in which a perfect matching exists.  In other words, it is possible to partition the triangles into a set of adjacent pairs (second column of Figure~\ref{fig:HamiltonianHorse}).  In practice, we use the Blossom-5 algorithm \cite{kolmogorov2009blossom} to find perfect matches of the dual graph.  Then, adding edges between unpaired triangles leads to a set of disconnected cycles (third column of Figure~\ref{fig:HamiltonianHorse}), which one can connect by a spanning set of edges between paired triangles.  Finally, the spanning edges are split into bridges between the disconnected cycles, joining them together into one large cycle that covers the surface of the triangle mesh.

\section{Curve Embedding in Audio}
\label{sec:methods}

We now introduce our new algorithm for hiding artistic space curves in audio.  Before we go into the details, we first define quantitative measurements for measuring the fit to the original audio (Goal~\ref{goal:imperceptible}) and the geometric quality of the hidden curve (Goal~\ref{goal:geomquality}).  Let the original carrier audio be $x$ and let the steganography audio be $y$, each with $N$ samples.  Then we define the steganographic signal to noise ratio in decibels (dB) as 

\begin{equation}
  \label{eq:stegsnr}
   \text{snr}(y|x) = 10  \log_{10} \left(\sum_{j=1}^N x_j^2 \right) -  10 \log_{10}\left(\sum_{j=1}^N (y_j-x_j)^2  \right)
\end{equation}

For our geometric measurement of quality, let $X_i$ be a sequence of $M$ target curve points in $\mathbb{R}^d$ and $Y_i$ be a sequence of $M$ reconstructed points in $\mathbb{R}^d$.  Then we define the {\em mean geometric distortion} is simply as the mean Euclidean distance between these points:
\begin{equation}
  \label{eq:distortion}
  \text{distortion}(Y|X) = \frac{1}{M} \sum_{j=1}^M ||X_j - Y_j||_2
\end{equation}

\subsection{Formulation of Least Squares Problem}
\label{sec:formulation}

We now formulate an objective function that trades off Equation~\ref{eq:stegsnr} and Equation~\ref{eq:distortion}.  Let $T = (T_1, T_2, \hdots, T_N )$ be the sequence of points of the target curve to hide, where each $T_i \in \mathbb{R}^d$, and let $T_{i, m}$ refer to $m^{\text{th}}$ coordinate of $T_i$, and let $x$ be a set of audio samples which will serve as a carrier.  The goal is to perturb the samples of $x$ so that some function of $x$ matches $T$ as closely as possible.  The function we choose is based on a ``time regularized'' version of the magnitude Short-Time Fourier Transform (STFT) which we call a {\em sliding window sum} STFT (SWS-STFT).  As a first step, we compute a {\em non-overlapping} STFT $S$ based on a chosen window length $w$ with a total of $N$ frames:

\begin{equation}
  S_{k, j} = \sum_{n = 0}^w x_{jw + n} \left(e^{-i 2 \pi k n / w} \right) = M_{k, j} \left( e^{i P_{k, j}} \right)
\end{equation}

and we factor $S$ it into its magnitude and phase components $M$ and $P$, respectively.  Next, we choose a subset of $d$ frequencies $k_i, i \in \{1, 2, \hdots, d\}$, each of which will hide a different dimension of $T$.  Given a second window length $\ell$, we then define the following {\em sliding window sum} function, which we apply to each row $k_i$ of the magnitudes of $S$ that we wish to perturb to obtain the SWS-STFT

\begin{equation}
  \text{SWS}^{\ell}(M)_{k_i, j} = \sum_{n = 0}^{\ell-1} M_{k_i, j+n}
\end{equation}

The effect of $\ell$ is to smooth out the noisy rows $k_i$ of the magnitude spectrogram so that the rows in the SW-STFT match smoother transitions in the target curves.  Each row of the SW-STFT has $N-\ell+1$ samples.  Let's assume momentarily that $T$ has exactly this many samples; we will address the case where $\text{length}(T) > N-\ell+1$ in Section~\ref{sec:reparam}.  We then seek a perturbed version of the magnitudes $\hat{M}$ so that each coordinate $i$ is hidden in a single frequency index $k_i$ of $\hat{M}$.  To that end, we minimize the following objective function, one coordinate dimension $i = 1, ... d$ at a time:

\begin{equation}
  \label{eq:objfn}
  f(\hat{M}_{k_i}) = \sum_{j=1}^{N-\ell+1} \left( \left( \sum_{n = 0}^{\ell-1} \hat{M}_{k_i, j+n} \right) - T_{i, j} \right)^2 + \lambda \sum_{j=1}^N \left( M_{k_i, j} - \hat{M}_{k_i, j} \right)^2
\end{equation}

\begin{figure}
  \centering
  \includegraphics[width=0.9\columnwidth]{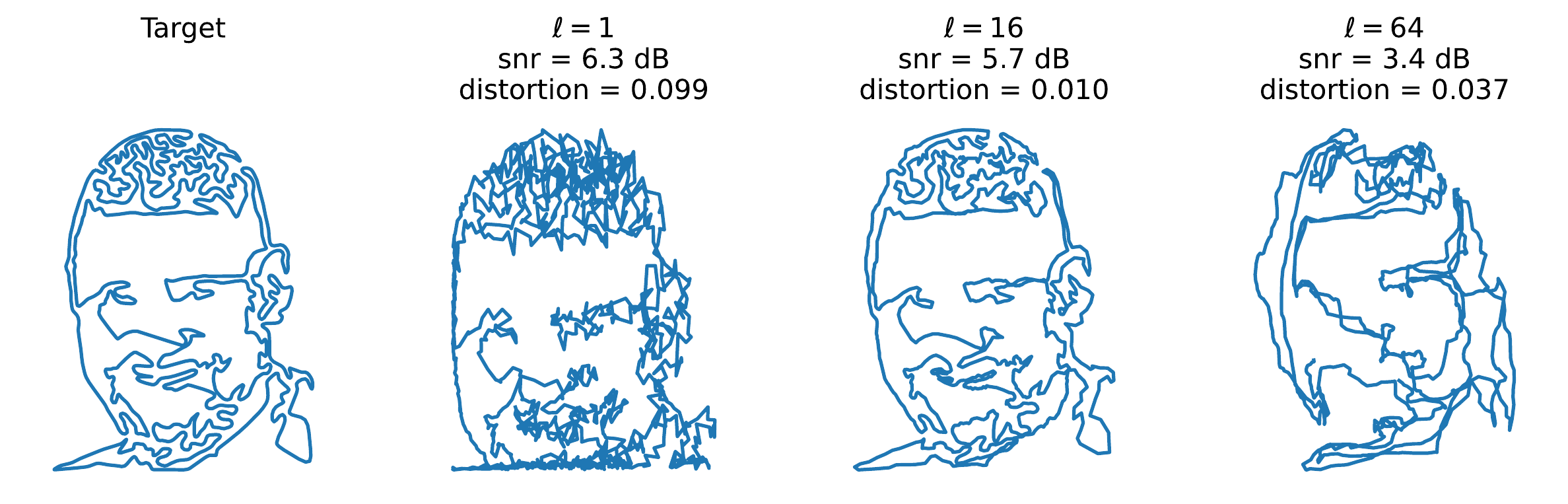}
  \caption{Varying $\ell$ for a fixed $\lambda=0.1$, using the lowest two non-DC frequencies to carry.  A larger $\ell$ for the SW-STFT in Equation~\ref{eq:objfn} leads to smoother curves which are more likely to survive compression, but an $\ell$ that's too large may over-smooth.}
  \label{fig:WindowEffect}
\end{figure}

\begin{figure}
  \centering
  \includegraphics[width=0.9\columnwidth]{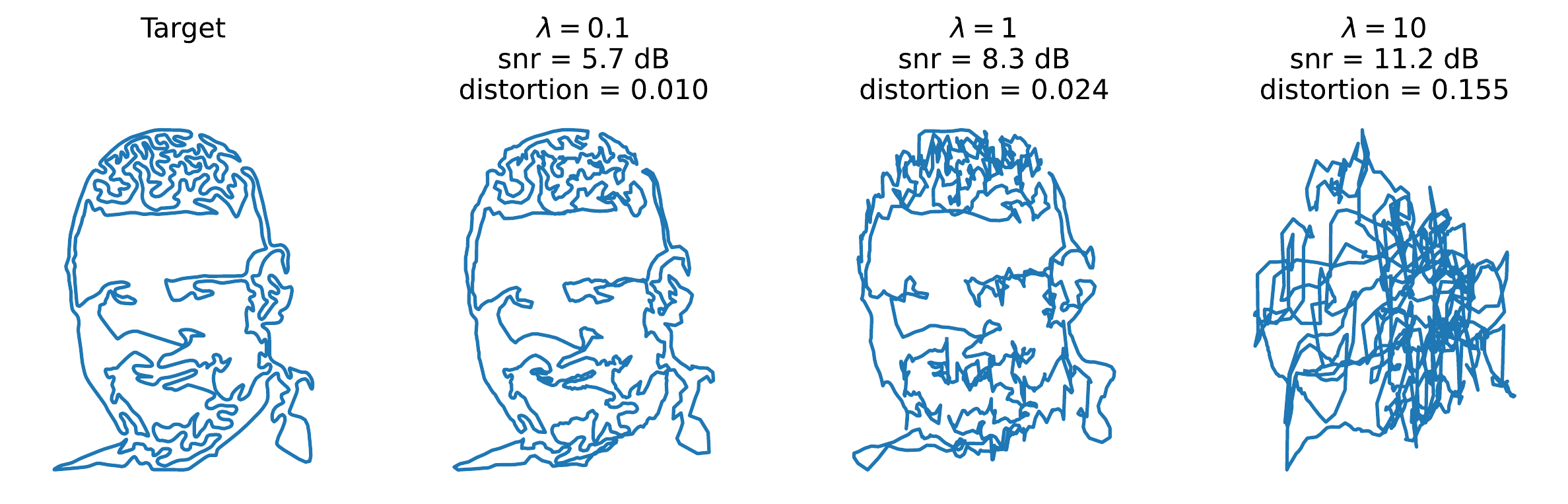}
  \caption{Varying $\lambda$ for a fixed $\ell=16$.  A smaller $\lambda$ in Equation~\ref{eq:objfn} leads to higher geometric fidelity (Goal~\ref{goal:geomquality}), at the cost of audio quality (Goal~\ref{goal:imperceptible}), as measured by SNR.}
  \label{fig:LambdaEffect}
\end{figure}

subject to $\hat{M_{k_i, j}} \geq 0$.  In other words, we want the magnitude SW-STFT of a perturbed signal to match the target coordinate as well as possible (minimizing Equation~\ref{eq:distortion}), while preserving the original audio as well as possible (minimizing Equation~\ref{eq:viterbiobj}), according to $\lambda$.  A greater $\lambda$ means that the signal $\hat{M}$ will fit the original audio better, at the cost of a noisier curve.  Figure~\ref{fig:LambdaEffect} shows an example.

After solving for $\hat{M_{k_i, j}}$, we replace all rows of $M_{k_i}$ with $\hat{M_{k_i}}$, and we perform an inverse STFT of $M e^{i P}$ using the original phases $P$ (Section~\ref{sec:componentscales} will explain when it is necessary to modify the phases as well).  We then save the resulting audio in a compressed format, and we recover the hidden signal by loading it and extracting the corresponding components of the magnitude SW-STFT.

\subsubsection{Computational Complexity}
\label{sec:computation}
Minimizing equation~\ref{eq:objfn} can be formulated as a sparse nonnegative linear least squares problem, and there are myriad algorithms (e.g. \cite{branch1999subspace}) for solving such systems efficiently via repeated evaluation of the linear system and its adjoint on iterative estimates of a solution .  Furthermore, though Equation~\ref{eq:objfn} suggest an $O(N \ell)$ time complexity to evaluate the objective function at each iteration, we implement the linear operator and its adjoint with $O(N)$ operations only, independent of $\ell$, using the 1D version of the ``summed area tables'' trick in computer vision \cite{lewisfast}.  For example, let $C_{k_i, j} = \sum_{i=0}^{j} \hat{M_{k, i}}$, and $C_{k_i, j < 1} = 0$.  Then Equation~\ref{eq:objfn} can be rewritten as 
\begin{equation}
  \label{eq:objfncumusum}
  f_i(\hat{M}) = \sum_{j=1}^{N-\ell+1} \left( (C_{k_i, j+\ell-1}-C_{k_i, j-1}) - T_{i, j} \right)^2 + \lambda \sum_{j=1}^N \left( M_{k_i, j} - \hat{M}_{k_i, j} \right)^2
\end{equation}

It is also worth noting that a higher $\lambda$ in equation~\ref{eq:objfn} leads to a lower condition number\footnote{The condition number of a matrix is defined as the ratio of the largest to smallest singular values, and lower condition numbers are more numerically desireable.} of the matrix in the implied linear system, which leads to faster minimization of Equation~\ref{eq:objfn}.  But as Figure~\ref{fig:LambdaEffect} shows, it may still be worth it to use smaller $\lambda$ values.  In practice, we see it as a difference between an encoding in 30 seconds of audio that takes a few seconds for $\lambda=0.1$ versus an encoding that takes a split second for $\lambda=10$ on a CPU.

\subsubsection{The choice of non-overlapping windows}
Though some other works use transforms with overlapping windows \cite{yun2009acoustic,geleta_pixinwav_2021}, changing different STFT bins independently leads to STFTs that do not correspond to any real signal due to discrepancies between overlapping windows.  An algorithm like Griffin-Lim \cite{griffin1984signal} can recover a signal whose spectrogram has a locally minimal distance to the perturbed STFT, but we find that this step introduces an unacceptable level of distortion both in the target and the carrier audio.  This problem occurs even for real-valued transforms such as the Discrete Cosine Transform (which was used by \cite{geleta_pixinwav_2021}).  Like \cite{xiaoxiao_dong_data_2004}, we sidestep this problem by using non-overlapping windows.  

Aside from window effects, the main downside of non-overlapping windows is that they lead to fewer STFT frames.  For instance, with a 30 second audio clip at 44100hz using a window length of 1024, we are limited to about 1292 frames, which is half of what we would get using an overlapping STFT with a hop length of 512.  However, a quick thought experiment shows that this is still a reasonable data rate.  Suppose that we use a sliding window length $\ell=16$, for a total of 1277 carrier samples.  Suppose also that the precision of our embedding of a 2D plane curve is roughly on par with an 8 bit per coordinate quantization in a more conventional binary encoding scheme.  Then the total number of bits transmitted over 30 seconds is $1277 \times 8 \times 2$, or about 681 bits/second.  This number jumps to about 1022 bits/second for a 3D curve.  These numbers are on par with other recent techniques that are designed to be robust to noise (e.g. 900 bits/second in the OFDM technique of \cite{eichelberger_receiving_2019}).  Furthermore, our equivalent of a ``bit error'' is additive coordinate noise, and the hidden signal degrades continuously with increased noise, rather than reaching a failure mode when a bit error is too high.

\subsection{Shifting, Scaling And Re-Parameterizing Targets for Better Fits}
\label{sec:reparam}

In this section, we explain how to modify the target curves to better match the given SW-STFT so that the STFT magnitudes don't have to be perturbed as much for the SW-STFT to match the target, leading to a less noticeable embedding of the hidden curve for the same quality geometry.

\subsubsection{Vertical Translation/Scaling}

A crucial step to keep the hidden signal imperceptible is to shift and rescale the target coordinates of the hidden curve to match the dynamic range of the SW-STFT components.  We first choose a scale ratio $a_i$ for each component as the ratio of the standard deviations $\text{stdev}_j \left( \text{SWS}^{\ell} (M)_{k_i, j} \right) / \text{stdev}_j (T_{i, j})$.  Then, letting $N$ be the length of the two signals, we compute the vertical shift $b_i$ as

\begin{equation}
  b_i = \frac{1}{N} \sum_{j=1}^N  \text{SWS}^{\ell} (M)_{k_i, j} - a_i T_{i, j}
\end{equation}

The rescaled target coordinate $\hat{T_i}$ is then defined as $\hat{T}_{i, j} = a_i T_{i, j} + b_i$.  Unfortunately, we lose relative scale information between the components, but we will explain how to hide and recover this in Section~\ref{sec:componentscales}

\subsubsection{Viterbi Target Re-Parameterization}

\algrenewcommand\algorithmicindent{0.8em}%
\begin{algorithm}
  \caption{Viterbi Target Re-Parameterization}

  \begin{algorithmic}[1]
    \Procedure{ViterbiTargetReparam}{$\hat{T}$, $\text{SWS}^{\ell} (M)$, $K$}
    \State $N_T \gets \text{len}(\hat{T})$ \Comment{Number of target points}
    \State $N_M \gets \text{len}(\text{SWS}^{\ell}(M))$ \Comment{Number of SW-STFT frames}
    \State $S[i>1, j] \gets 0$ \Comment{$N_T \times N_M$ Cumulative cost matrix}
    \State $S[1, j] \gets \sum_{i=1}^d (\hat{T}_{i, 1} - \text{SWS}^{\ell} (M)_{k_i, j})^2$
    \State $B[i, j] \gets 0$ \Comment{$N_T \times N_M$ backpointers to best preceding states}
    \For{$j = 2:N_M$}
        \For{$t = 1:N_T$}
            \State $S[t, j] = \min_{k=(t-K) \mod N_T} ^ {k=(t-1) \mod N_T} S[k, j-1] $ \Comment {Find best preceding state}
            \State $S[t, j] \gets S[t, j] + \sum_{i=1}^d (\hat{T}_{i, t} - \text{SWS}^{\ell} (M)_{k_i, j})^2$ \Comment{Add on matching cost}
            \State $B[t, j] = \argmin_{k=(t-K) \mod N_T} ^ {k=(t-1) \mod N_T} S[k, j-1] $ \Comment {Save reference to best preceding state}
        \EndFor
    \EndFor
    \State Backtrace $B$ to obtain the optimal sequence $\Theta$ \\
    \Return $\Theta$
    \EndProcedure
  \end{algorithmic}
  \label{alg:viterbiwarp}
\end{algorithm}

Beyond shifting and scaling the targets coordinates vertically, we may also need to re-parameterize them in time, since, in general, the SWS-STFT sequence will not have the same number of samples as the shifted target $\hat{T}$.  Furthermore, the target curves are cyclic, so the starting point is arbitrary, nor does it matter if we traverse the curve left or right, or at a constant speed.  This gives us a lot of freedom in how we choose to re-parameterize the target to best match the signal even before we perturb the signal.  Let $N_M$ be the number of frames in the SW-STFT and $N_T$ be the number of samples of the target, and assume that $N_T > N_M$ (we can always resample our target curves to make this true).  Let $\Theta = \{ \theta_1, \theta_2, \theta_3, \hdots, \theta_{N_M} \}$ be new indices into the target, and let $K > 0$ be a positive integer so that $1 \leq \left( (\theta_i - \theta_{i-1}) \mod N_T \right) \leq K$; that is, $K$ is the maximum number of samples by which the target re-parameterization can jump in adjacent time steps.  We seek a $\Theta$ minimizing the following objective function for a $d$-dimensional shifted target $\hat{T}$

\begin{equation}
  \label{eq:viterbiobj}
  g(\Theta = \{\theta_1, \theta_2, \hdots, \theta_{N_M}\}) = \sum_{j = 1}^{N_M} \sum_{i=1}^d \left( \text{SWS}^{\ell} (M)_{k_i, j} - \hat{T}_{i, j}  \right)^2
\end{equation}

For a fixed $K$, we use the Viterbi algorithm to obtain a $\Theta$ minimizing Equation~\ref{eq:viterbiobj} in $O(MNK)$ time.  Algorithm~\ref{alg:viterbiwarp} gives more details.  

\begin{figure}
  \centering
  \includegraphics[width=\columnwidth]{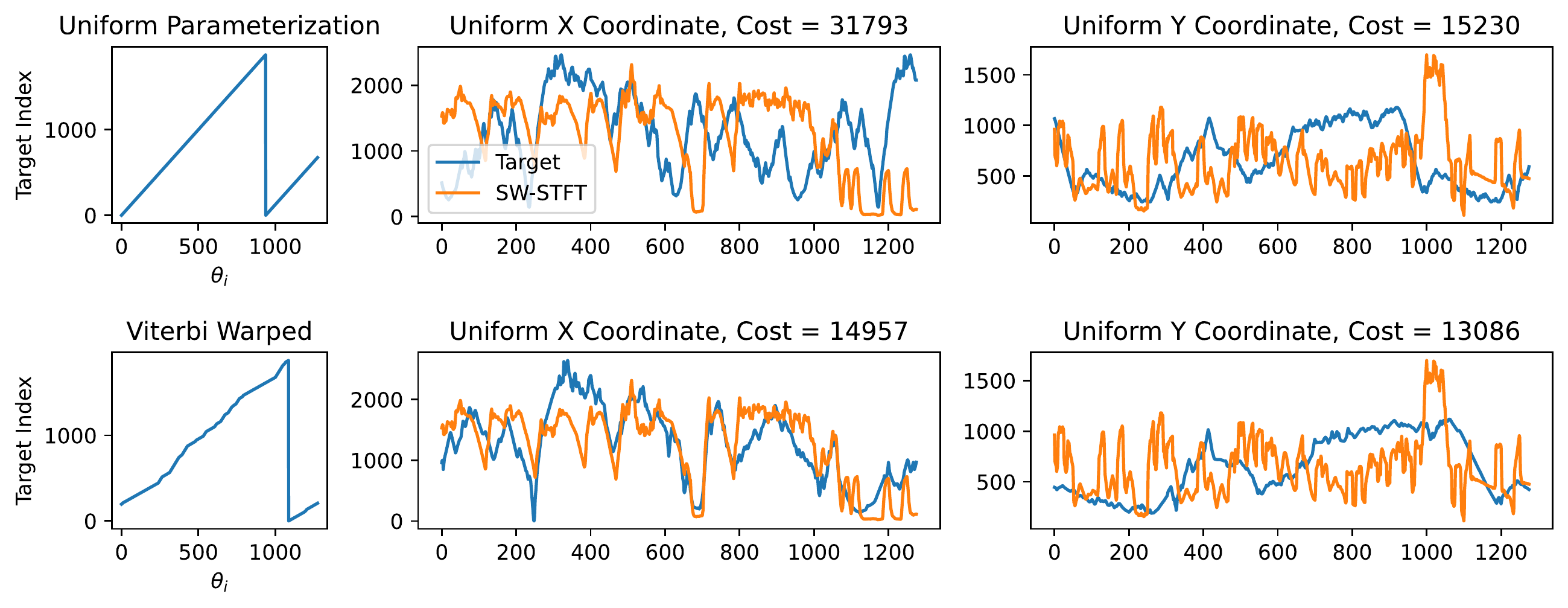}
  \caption{Circularly shifting the target loop and traversing it at a non-uniform speed traces out the same shape, while matching the given SW-STFT better than a uniform parameterization starting at an artibrary place on the target.}
  \label{fig:ViterbiWarp}
\end{figure}

In practice, we re-run the algorithm starting at $K = 1$, and we repeatedly increment $K$ until the optimal $\Theta$ goes through at least one full loop on the target.  Since clockwise or counter-clockwise traversal of the target is arbitrary, we then rerun this procedure again for a reversed version of the sequence and keep the result which minimizes Equation~\ref{eq:viterbiobj}.  As a rule of thumb, we find that having a target curve with about 1.5-2x as many samples as there are SWS-STFT frames gives enough wiggle room for the Viterbi algorithm.  In our experiments in Section~\ref{sec:experiments}, we will generate TSP and Hamiltonian sequences with 2000 samples for our 1200-1300 SWS-STFT frames.  Figure~\ref{fig:ViterbiWarp} shows an example running this algorithm under these conditions on the Usher example in column 3 of Figure~\ref{fig:WindowEffect}.

\subsection{Storing Component Scales in Phase}
\label{sec:componentscales}

Since we intentionally rescale the dimensions of the target to match the dynamic range of each SW-STFT component, we lose the aspect ratio between the dimensions.  But since we have only perturbed the magnitude components of the frequency indices $k_i$, we still have some freedom to perturb the phases to store additional information.  To that end, we use the technique presented by the authors of \cite{xiaoxiao_dong_data_2004} to store the relative scale of each dimension in the phase.  We store the same scale in the phase of every STFT frame, and we take the scale to be the median of the phases upon decoding.

\subsection{Recovering Frame Alignments}
\label{sec:framealignments}

\begin{figure}
  \centering
  \includegraphics[width=\columnwidth]{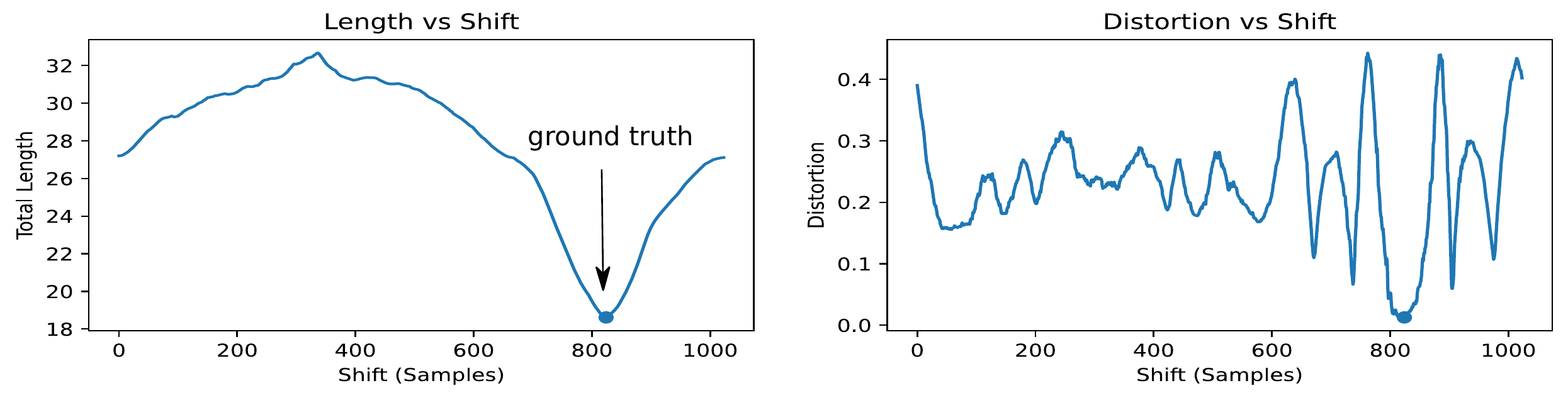}
  \caption{A shift which minimizes curve length is most likely the shift needed to re-align audio to frame windows.  In this example with the embedding using the lowest two non-DC frequencies with $\lambda=0.1, \ell=16$, the global mins exactly match ground truth.}
  \label{fig:FrameAlignments}
\end{figure}

What we've described so far works for audio that is aligned to each window, but additional work needs to be done to address Goal ~\ref{goal:misalignment} if the audio to decode comes in misaligned.  To this end, we use the fact that the hidden curves move only slightly between adjacent samples; a TSP tour is defined as length-minimizing, and adjacent samples in Hamiltonian cycles on meshes move only between neighboring triangles on the original mesh.  If the embedding is frame aligned, the length of the curve should be minimized.  Conversely, if the embedding is not frame aligned, the curve becomes noisy and is more likely to jump around quickly from sample to sample.  Therefore, we can pick the alignment which minimizes the length in all possible shifts from $0$ to the STFT window length.  Figure~\ref{fig:FrameAlignments} shows an example.  We will empirically evaluate this in Section~\ref{sec:experiments}.

\section{Experiments}
\label{sec:experiments}

\begin{figure}
  \centering
  \includegraphics[width=\columnwidth]{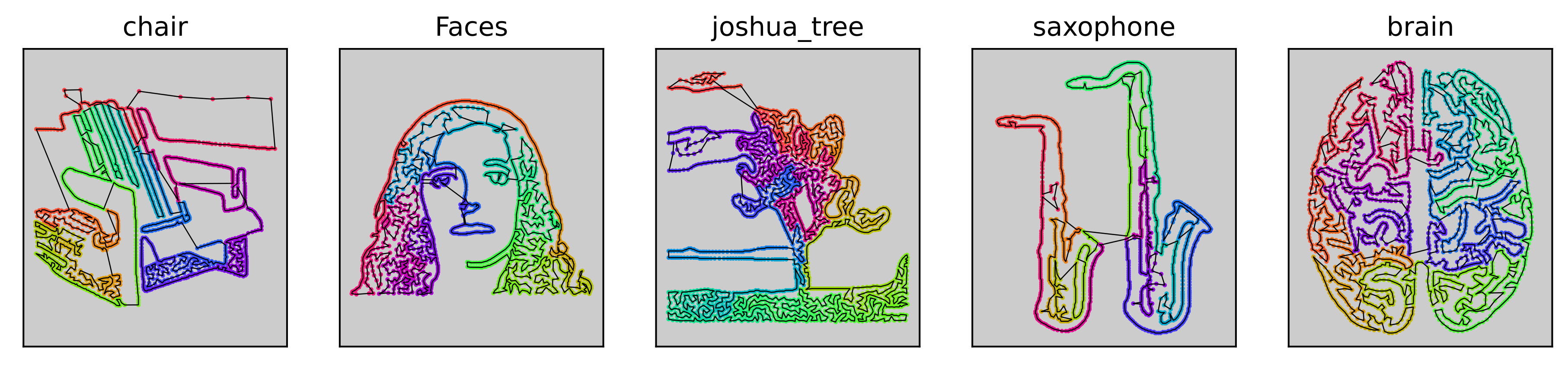}
  \caption{Examples of TSP art on the Caltech-101 dataset \cite{li_andreeto_ranzato_perona_2022}.}
  \label{fig:caltech101examples}
\end{figure}

\begin{figure}
  \centering
  \includegraphics[width=\columnwidth]{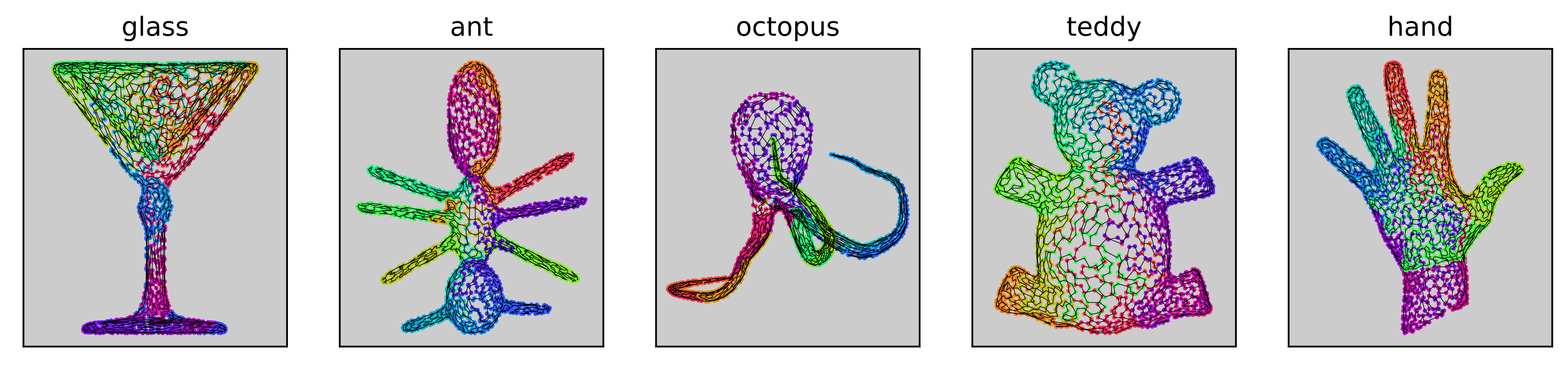}
  \caption{Examples of Hamiltonian cycles on triangle meshes from the Princeton mesh segmentation benchmark \cite{Chen:2009:ABF}.}
  \label{fig:meshsegexamples}
\end{figure}

We now quantitatively assess the performance of our system.  To generate a large set of curves, we generate TSP art on the roughly 10,000 images in the Caltech-101 dataset  \cite{li_andreeto_ranzato_perona_2022} (e.g. Figure~\ref{fig:caltech101examples}), and we generate Hamiltonian cycles on the 380 watertight triangle meshes in the Princeton mesh segmentation database \cite{Chen:2009:ABF} (e.g. Figure~\ref{fig:meshsegexamples}).  We then use the 1000 30 second audio clips from the Tzanetakis genre dataset \cite{tzanetakis2002musical} as carrier audio.  For the Caltech-101 database, we partition the images into sets of 10, which are each encoded in one of the audio carriers varying $\lambda$ and $\ell$.  Likewise, for the mesh segmentation dataset, we hide each Hamiltonian path in three different audio clips from the Tzanetakis dataset.  In all cases, we use an STFT window length of 1024 at a sample rate of 44100hz, and we encode the audio using lossy mp3 compression at 64 kbps.

\begin{figure}
  \centering
  \includegraphics[width=\columnwidth]{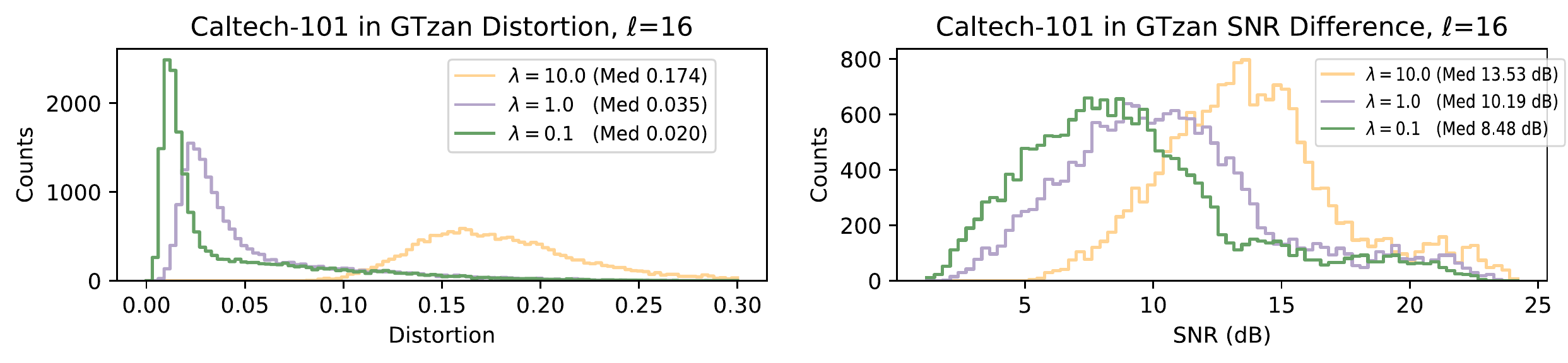}
  \includegraphics[width=\columnwidth]{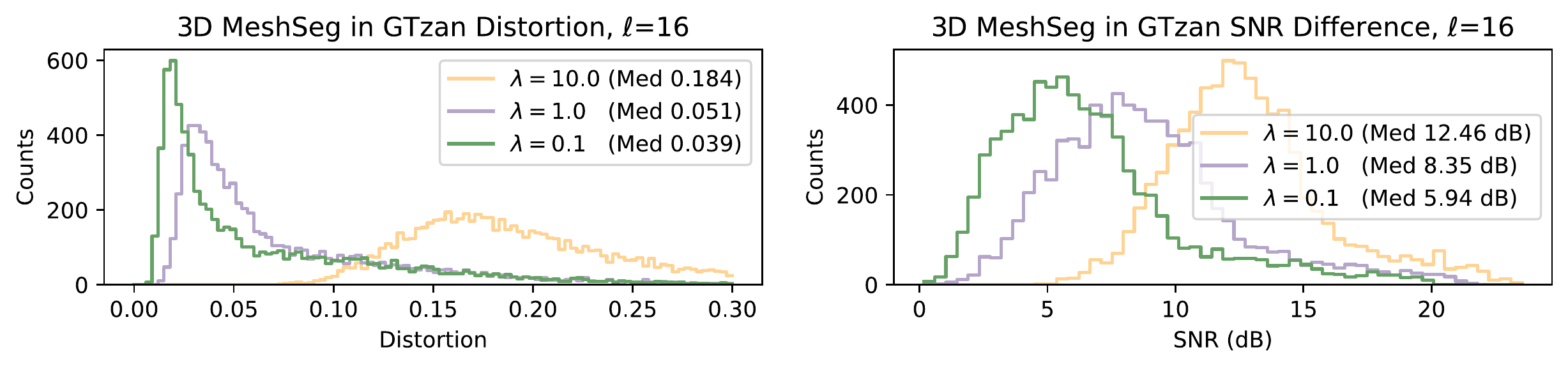}
  \caption{The results embedding curves into clips from the Tzanetakis genre dataset, for a fixed $\ell=16$.  As expected from Equation~\ref{eq:objfn}, both the distortion and SNR go up as $\lambda$ increases.}
  \label{fig:ResultsFixedWin}
\end{figure}

\begin{figure}
  \centering
  \includegraphics[width=\columnwidth]{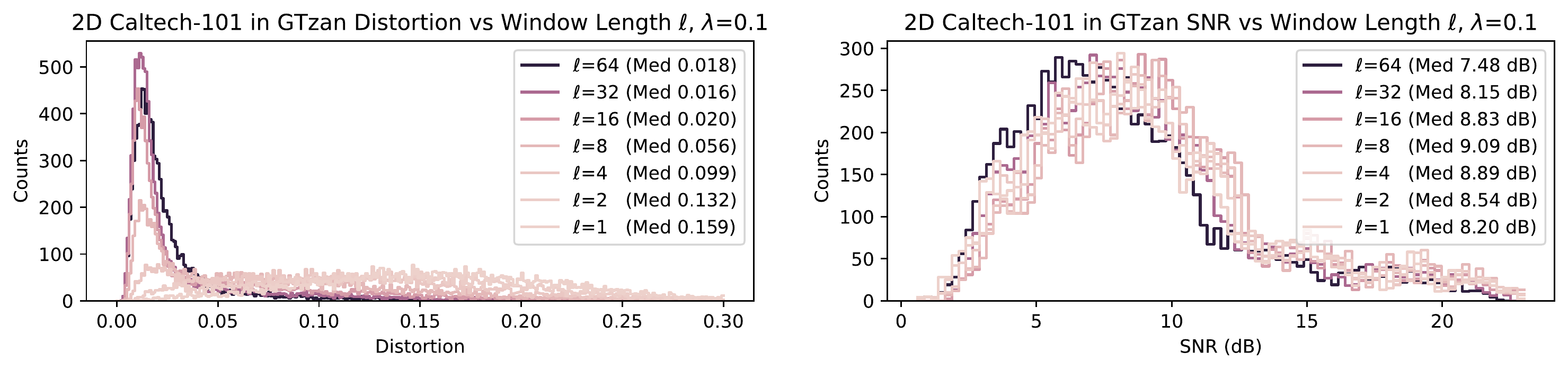}
  \caption{Embedding curves into clips from the Tzanetakis genre dataset varying the window length $\ell$ for a fixed $\lambda$.  Moderate window lengths are the best choices for both SNR and distortion.  We recommend $\ell=16$. (3D is similar; see supplementary)}
  \label{fig:ResultsFixedLam}
\end{figure}

\begin{figure}
  \centering
  \includegraphics[width=\columnwidth]{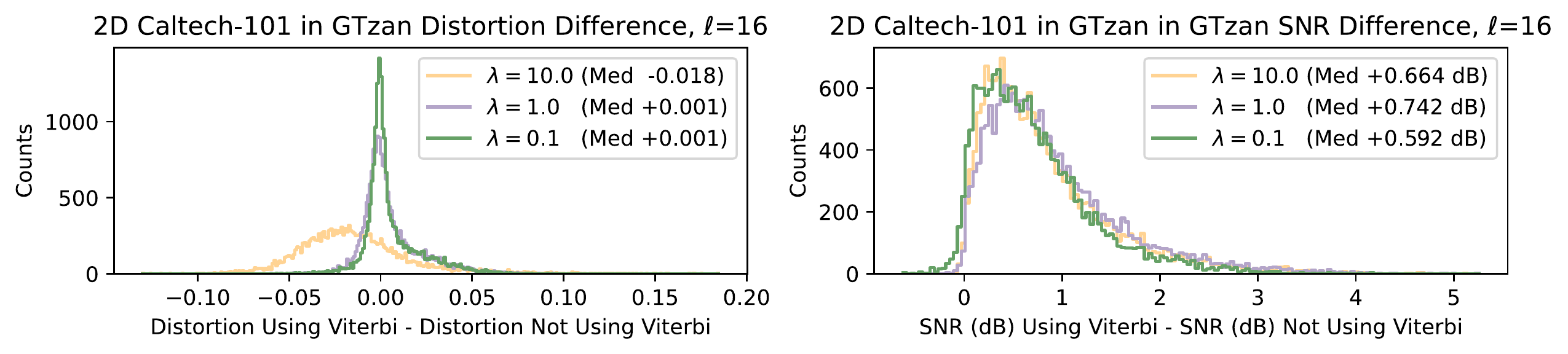}
  \includegraphics[width=\columnwidth]{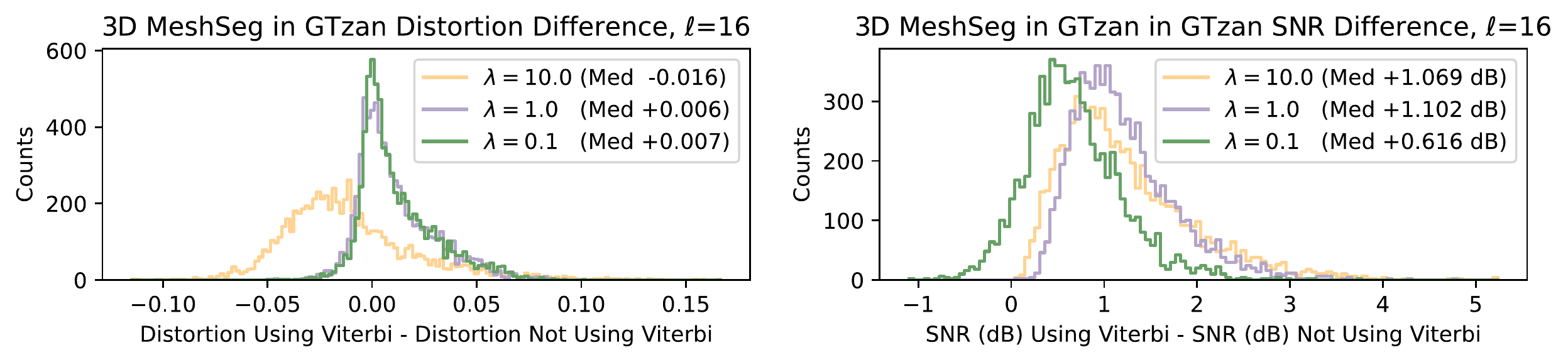}
  \caption{Pre-warping the target with Viterbi alignment (Section~\ref{sec:reparam}) overall improves the resulting distortion and SNR.}
  \label{fig:ResultsViterbiExperiment}
\end{figure}

\begin{figure}
  \centering
  \includegraphics[width=\columnwidth]{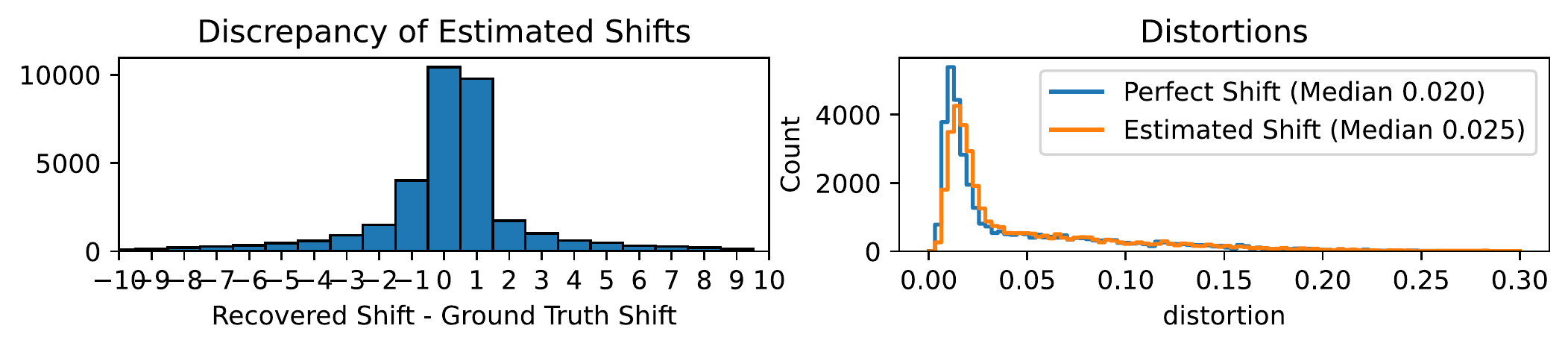}
  \caption{Estimating the frame alignment by minimizing the length (Section~\ref{sec:framealignments}) works nearly as well as perfect knowledge of the alignment. Using $\lambda=0.1, \ell=10$ over all images in Caltech-101, 93\% of the shifts are within 10 of the ground truth for an STFT window length of 1024, which has hardly any effect on the geometry of the curve.}
  \label{fig:ShiftsExperiment}
\end{figure}

Overall, we see slightly higher distortions and lower SNRs for 3D embeddings than 2D embeddings, which makes sense since there is one additional coordinate to hide in 3D.  As expected, increasing $\lambda$ increases both the SNR and distortion, as shown in Figure~\ref{fig:ResultsFixedWin}. Also, as Figure~\ref{fig:ResultsFixedLam} shows, increasing the window length has a positive effect on geometric distortion, while moderate window lengths lead to the best SNR.  We also see a positive effect of the Viterbi alignment from Section~\ref{sec:reparam} on the SNR in Figure~\ref{fig:ResultsViterbiExperiment}.  Though $\approx$+1dB may not seem significant, it can make a huge difference in audio, particularly in quiet regions.  Finally, we run an experiment on the Caltech-101 dataset by choosing 4 random frame offsets per embedding, and we see in Figure~\ref{fig:ShiftsExperiment} that the algorithm of Section~\ref{sec:framealignments} recovers alignments well.

\subsubsection{Subjective Listening Experiment}

\begin{figure}
  \begin{minipage}[c]{0.36\textwidth}
    \caption{
      Results of the listening experiment on the Amazon Mechanical Turk.  A lower $\lambda$ leads to a lower mean opinion score, as expected, though not to an intolerable degree.
    } \label{fig:TurkResults}
  \end{minipage}
  \begin{minipage}[c]{0.64\textwidth}
    \includegraphics[width=\columnwidth]{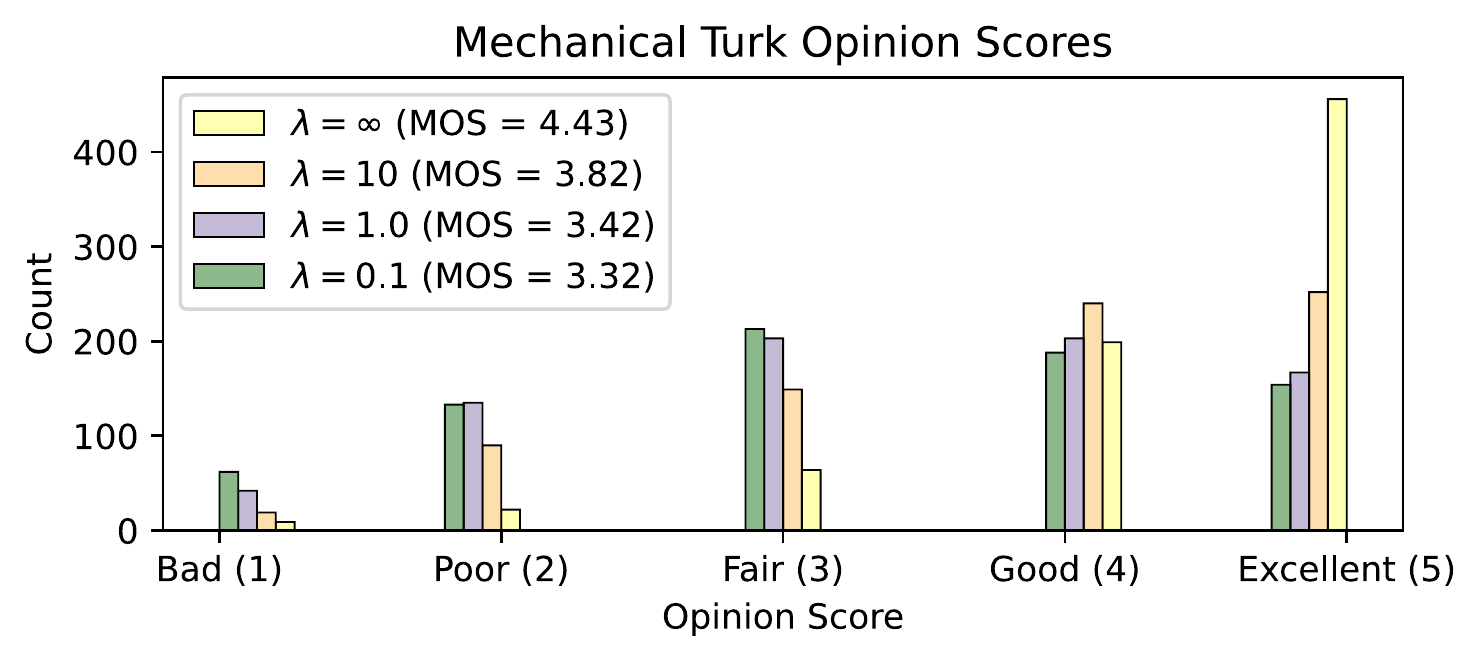}
  \end{minipage}\hfill
  
\end{figure}

Though the experiments above are encouraging, SNR can be misleading; frequencies that are more audible may actually have a higher SNR due to psychoacoustic phenomena.  To address this, we performed a crowd-sourced listening experiment on the Amazon Mechanical Turk where we embedded a random image from the Caltech-101 dataset in each of the Tzanetakis clips.  We split them into 4 groups with no embedding (control), and with $\lambda = 0.1, 1, 10$.  We asked the listeners to rate the quality of the noise on the 5 point impairment scale of \cite{bassia2001robust} (5: imperceptible, 4: perceptible but not annoying, 3: slightly annoying, 2: annoying, 1: very annoying).  In our experiment, we had 46 unique Turkers, 21 of whom participated in at least 40 rankings.  Figure~\ref{fig:TurkResults} shows the results.  Mean opinion scores (MOS) are correlated with $\lambda$, but there is little difference between $\lambda=0.1$ and $\lambda=1$, which suggests using the former as a rule of thumb due to its lower geometric distortion.

\section{Discussion / Supplementary Material}

We have presented a model-based scheme for hiding artistic curves in audio, and the curves survive lossy compression while remaining reasonably imperceptible, as validated both with quantitative measurements and by humans.  We hide the dimensions of curves in time regularized magnitudes of STFT frequencies, though coefficients of any orthogonal decomposition could work (we also implemented wavelets, though we found them more audibly perceptible).  Our scheme is incredibly simple and requires no training.  Decoding is nearly instantaneous, as it only requires computing the STFT of a few frequencies. 

To show off our pipeline, we created an interactive viewer in Javascript using WebGL that can load in and decode any mp3 file.  The viewer plays the decoded curve synchronized to the music.  We provide a variety of 2D and 3D precomputed examples to demonstrate our capabilities.  To view our supplementary material, source code for encoding and decoding, and live examples, please visit \url{https://github.com/ctralie/AudioCurveSteganography}


%
%
%
\bibliographystyle{splncs04}
\bibliography{tralie}

\end{document}